\documentclass[journal]{IEEEtran}

\usepackage{caption,subcaption,float}

\usepackage{multirow}
\usepackage{graphicx}
\usepackage{multirow}
\usepackage{amsmath,fixmath,amsfonts}
\usepackage{amssymb}
\usepackage{array}
\usepackage{ifpdf}
\usepackage{cite}
\usepackage{url}
\usepackage{caption} 
\usepackage{hyperref} 
\begin{document}
\title{A Versatile Depth Video Encoding Scheme Based
on Low-rank Tensor Modeling for Free Viewpoint
Video}

\author{Mansi~Sharma,
        Jyotsana Grover
\thanks{M. Sharma is Computer Science and Engineering with Thapar Institute of Engineering and Technology, Patiala, Punjab 147004 and J. Grover is with Computer Science \& Information Systems, Birla Institute of Technology and Science, Pilani, Pilani Campus

E-mail: \{mansi.sharma@thapar.edu,~jyotsana.grover@pilani.bits-pilani.ac.in\}.
}}

\maketitle

\begin{abstract}
The compression quality losses of depth sequences determine quality 
of view synthesis in free-viewpoint video. The depth map intra prediction in 
3D extensions of the HEVC applies intra modes with auxiliary depth modeling modes (DMMs) to better preserve depth edges and handle motion discontinuities. Although such modes enable high efficiency compression, but at the cost of very high encoding complexity. Skipping conventional intra coding modes and DMMs in depth coding limits practical applicability of the HEVC for 3D display applications. In this paper, we introduce a novel low-complexity scheme for depth video compression based on low-rank tensor decomposition and HEVC intra coding. The proposed scheme leverages spatial and temporal redundancy by compactly representing the depth sequence as a high-order tensor. Tensor factorization into a set of factor matrices following CANDECOMP/PARAFAC (CP) decomposition via alternating least squares give a low-rank approximation of the scene geometry. Further, compression of factor matrices with HEVC intra prediction support arbitrary target accuracy by flexible adjustment of bitrate, varying tensor decomposition ranks and quantization parameters. The results demonstrate proposed approach achieves significant rate gains by efficiently compressing depth planes in low-rank approximated representation. The proposed algorithm is applied to encode depth maps of benchmark ``Ballet'' and ``Breakdancing'' sequences. The decoded depth sequences are used for view synthesis in a multi-view video system, maintaining appropriate rendering quality. 
\end{abstract}

\IEEEpeerreviewmaketitle

\section{Introduction}
\label{sec:intro}
The market of 3D videos will continue to flourish in upcoming days.
The strongest demand of 3D contents currently comes from industries in the creative economy and immersive media such as Gaming, Virtual Reality (VR) or Augmented Reality (AR) platforms, Displays, Film Entertainment, Imaging Systems and Retail. More pervasive applications of 3D videos, we will find in industries as diverse as healthcare, education, military, and the real estate over time. The advancement of affordable Time-of-flight (ToF) sensing or 3D depth cameras, further, promote application of 3D videos in academia and industry. There exist a variety of 3D video content formats in the user mass market to support 3D applications \cite{IntroRefnew1}. This includes conventional stereo video (CSV), mixed resolution stereo (MRS), hybrid mono-stereo, video plus depth (V+D), multiview video (MVV), multiview video plus depth (MVD), and layered depth video (LDV). These formats are specifically used for rendering in 3D and VR head mounted displays \cite{IntroRef1,IntroRef2}. Moving Picture Experts Group (MPEG) standardized MVD and LDV representations and depth-image-based rendering (DIBR) for flexible display adaptation of current 3D display technology. Latest call for 3D video standardization support depth enhanced stereo (DES) as a generic backward compatible 3D video format that would support extended functionality like baseline adaptation, post production, multi-view rendering, specific to display types and sizes \cite{IntroRef4,IntroRef5}. The objective is to decouple content creation from the display system requirement. The key functionality in this context is identified as manipulation of the depth composition of a scene via view synthesis \cite{IntroRef6,IntroRef7}. Owing to the constraints in transmission and broadcasting channels, only limited texture views and their corresponding depth maps are acquired, compressed, and transmitted for 3D video applications. At decoder side, with the aid of depth information, DIBR is employed to synthesize an arbitrary number of views to support head motion parallax viewing and baseline adaptation within practical limits of different multiview display types. The perceptual quality of synthesized novel views is mainly determined by the quality of RGB images and depth maps as well as DIBR algorithm \cite{IntroRef8,IntroRef9}. Predominately, rendered view quality is sensitive to the compression distortion in depth images.

\begin{figure*}[t]
\centering
\includegraphics[height=6cm,width=15cm]{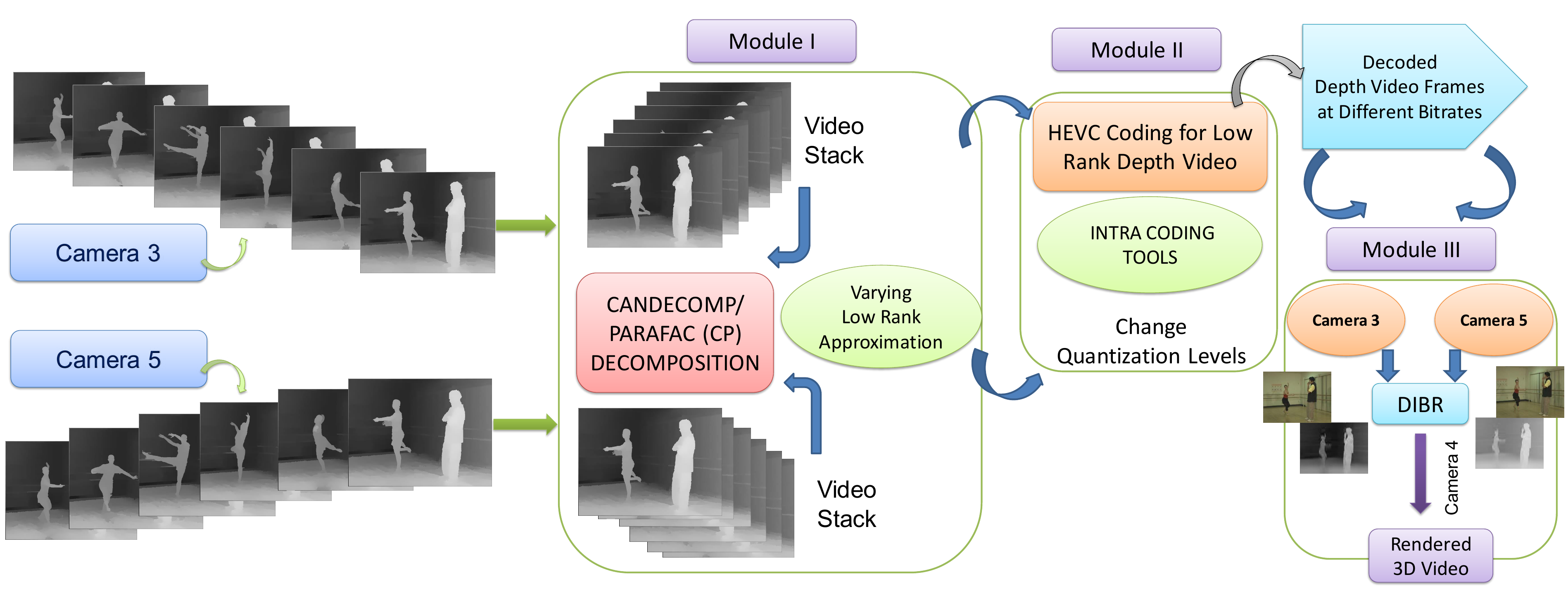}
\caption{Workflow of depth video compression scheme for 3D displays.}
\label{Fig1}
\end{figure*}
A depth map is basically stored as a gray-level image. It describes 
the relative distance from recording camera to actual objects
in the 3D space \cite{Ref1}. The characteristics of depth maps are fairly different from RGB texture images. Typically, a depth map consists of smooth regions and sharp edges around the object's boundaries \cite{IntroRef10}.
The compression distortion of sharp edges causes visual artifacts such as ringing effect at object's boundaries in synthesized views using DIBR algorithm  \cite{IntroRef8,IntroRef9}. Thus, preserving sharp edges is a critical task for depth video coding and high-quality view synthesis. The conventional 2D block matching based video coding algorithms are suboptimal for depth image sequences \cite{IntroRef10,IntroRef11,IntroRef12}. Such approaches generally divide large homogeneous regions in depth maps into small blocks. The sharp discontinuities around object boundaries can be placed within the same block. Block matching with arbitrary motion prediction is fairly demanding. This leads to significant coding artifacts around the depth edges in decoded depth images at high compression ratios \cite{Ref10, IntroRef13, IntroRef14, IntroRef15}.

\begin{figure*}[t]
\centering
\begin{tabular}{cccc}
\subfloat{\includegraphics[width = 3.0in]{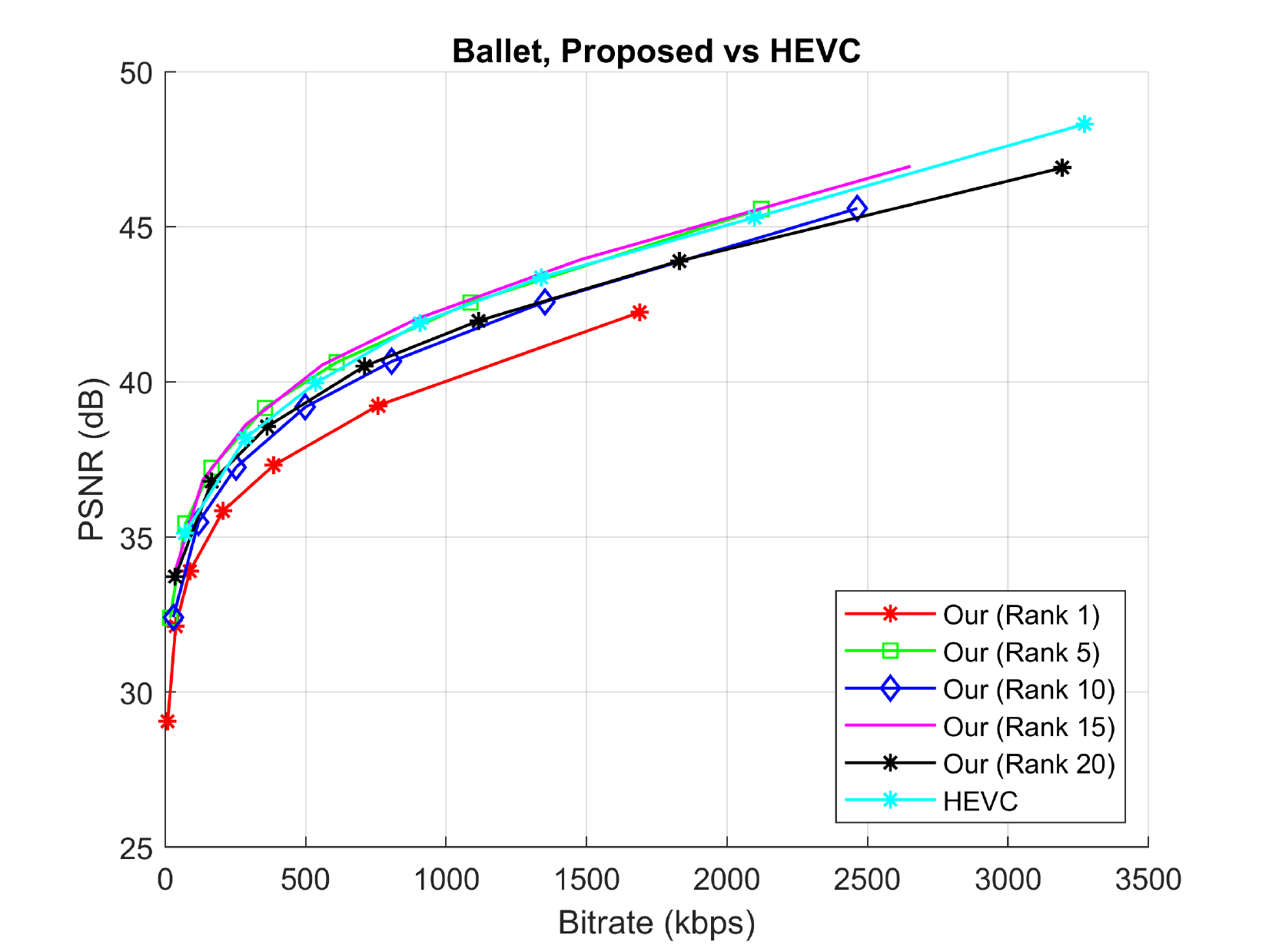}}
\subfloat{\includegraphics[width = 3.0in]{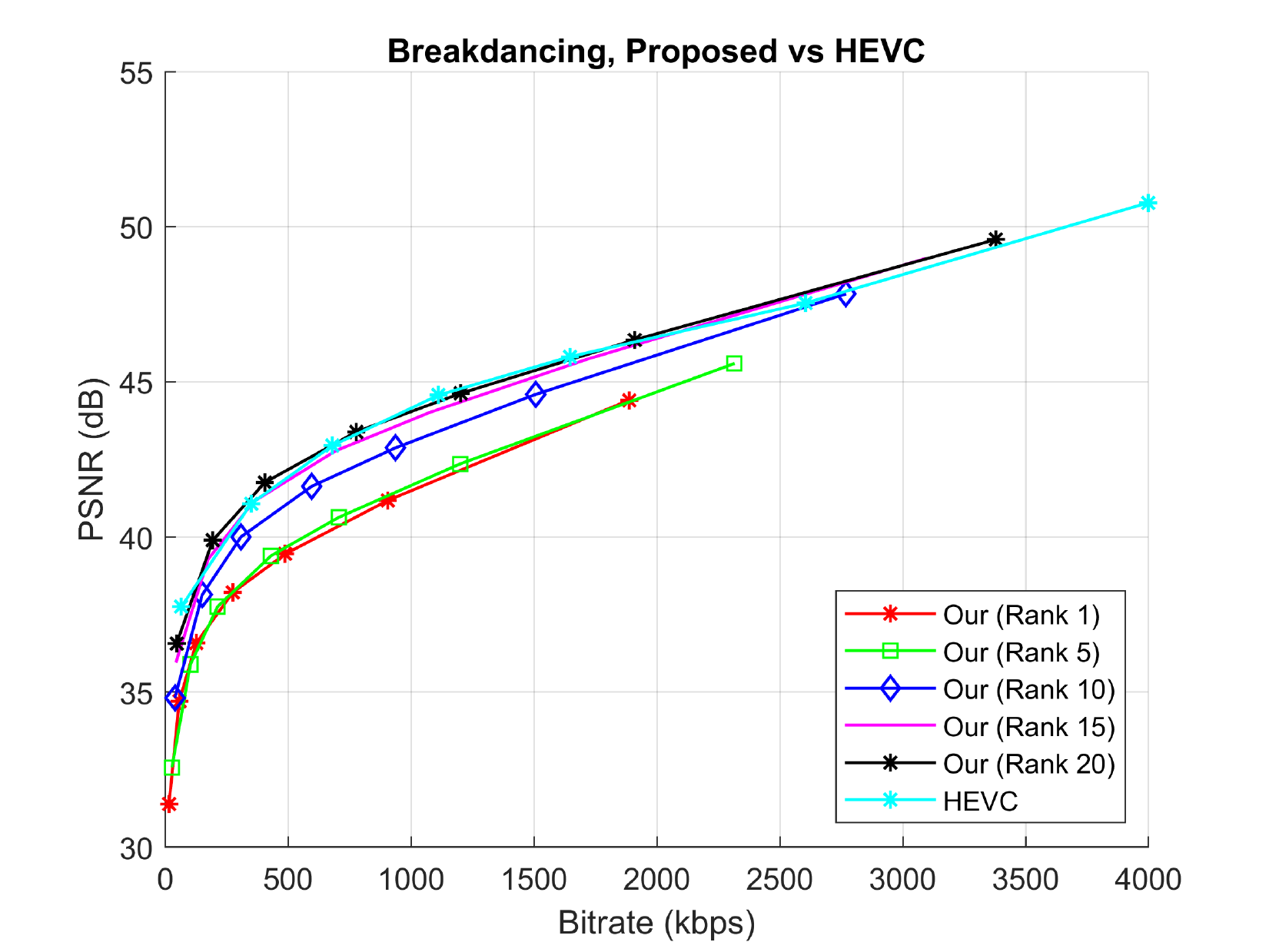}}\\
\subfloat{\includegraphics[width = 3.0in]{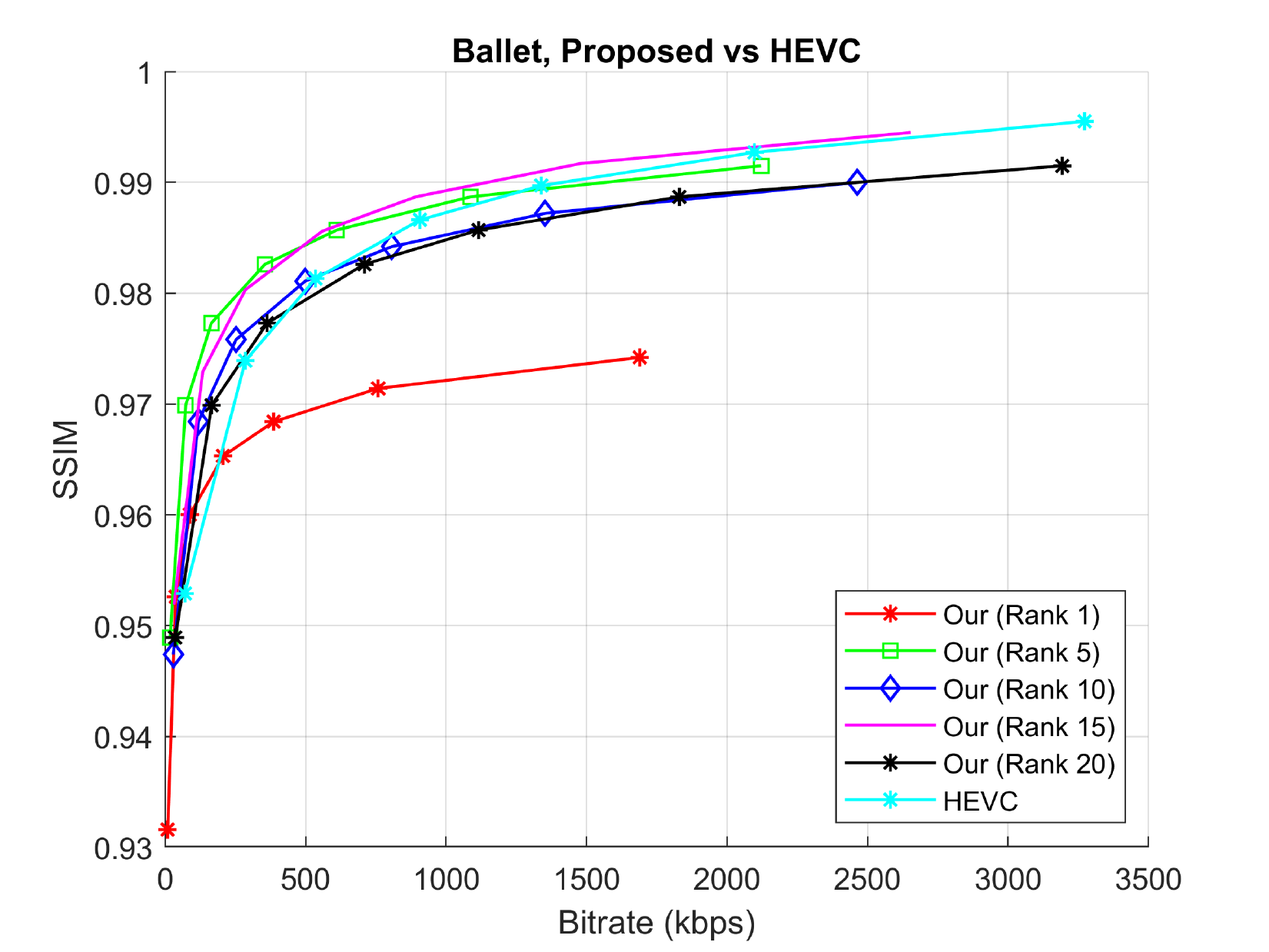}}
\subfloat{\includegraphics[width = 3.0in]{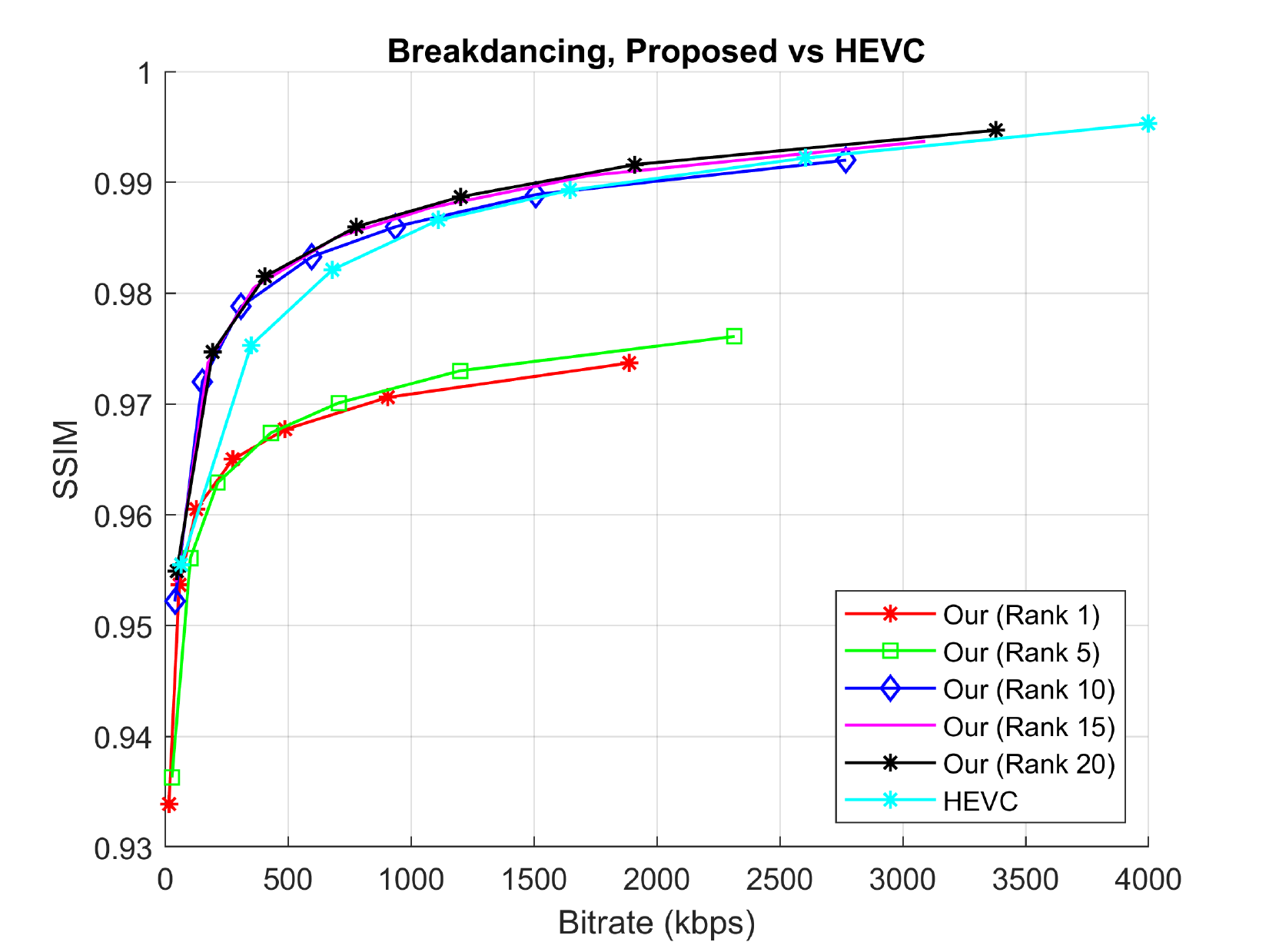}}
\end{tabular}
\caption{Rate distortion (RD) curve comparison between the proposed coding scheme and HEVC codec; x-axis: bitrate to code left (camera 3) and right camera (camera 5) depth sequences, y-axis: PSNR and SSIM scores of the synthesized intermediate camera view (camera 4).
}
\label{Fig2}
\end{figure*}

Joint Collaborative Team on Video Coding (JCT-VC) introduced a new generation of coding standards for 3D video such as High Efficiency Video Coding (HEVC) \cite{IntroRef17}, 3D-HEVC \cite{IntroRef18}, MV-HEVC \cite{IntroRef19}, etc. New tools have been introduced such as depth modeling mode (DMM) \cite{IntroRef20}, segment-based depth coding (SDC) \cite{IntroRef21}, and view synthesis optimization (VSO) \cite{IntroRef20,IntroRef22,IntroRef23} to explicitly handle depth video coding with consideration of preserving sharp edges. DMM intra mode of depth maps in 3D-HEVC offers flexibility in representing the sharp edges by following a non-rectangle partition approach based on wedgelet \cite{IntroRef24} or contour \cite{IntroRef25} based segmentation. It can save about 5\% of the transmitted bitrate maintaining acceptable view quality. The SDC approach chooses an alternative coding path and compresses the residual signal with constant pixel value, instead of following conventional transformed and quantization coefficients. The VSO mode selects different coding modes and the unit partition to trade off rate-distortion (RD) optimization and synthesized view distortion. In VSO evaluation, all combinations of block sizes need to be checked, considering DMM
intra mode with and without SDC. This improves the coding efficiency, but at the expense of high computational complexity. This issue cannot be solved even adopting complex HEVC quadtree coding structure in 3D-HEVC \cite{3DHEVCRef1}.

Several other research efforts are made to remove temporal 
or intra/inter-view redundancies in depth video images considering 
the specific properties of depth images \cite{MERef1,MERef2}. The basis of such schemes is motion estimation considering 2D block matching algorithms. Most of the algorithms exploit correlation between the RGB color and depth frames. Grewatsch and Miiller \cite{MERef1} adopts a conventional block matching approach and determine motion vectors (MVs) using the RGB video. The estimated MVs from texture information are then considered for encoding both the texture and depth sequences. Contrary, Daribo et al. \cite{MERef2} considered the motion of a block at the same coordinates in both texture and depth videos and adopted a joint distortion criterion to estimate common MVs for both texture and depth frames. Such approaches remove temporal or inter-view redundancies if accurate estimation of motion can be determined. Thus, these algorithms work well for coding the depth images acquired from cameras that remain in a fixed position. The case becomes inevitably much more complicated when moving cameras are involved, such as handheld RGB-D capturing. Unlike static camera acquisition, the same objects depth values change in successive depth frames as the distance between a mobile camera and the objects in a scene change across the time. Therefore, motion compensation based depth coding schemes output very inaccurate results with mobile RGB-D cameras 
\cite{IntroRef14, IntroRef15, Ref6, Ref10}. 

The above mentioned depth video codecs/methods usually can only work for a specific bitrate. This restriction of one system per bps (bits per second) limits the generality and flexibility of existing codecs/methods for practical 3D video rendering applications. This paper proposed a novel coding scheme which flexibly performs depth video compression at multiple bitrates, accomplishing varying quality levels. The idea of lossy compression is inspired from low-rank modelling by tensor approximation that represents the high-dimensional depth data more compactly. A simple yet effective CANDECOMP/PARAFAC (CP) tensor decomposition based scheme is developed, where there is a tensor  modelling block to approximate a low-rank representation of depth video data into a set of factor matrices. By differing the rank of factor matrices in tensor decomposition and quantization parameters of the HEVC intra coding block, one can flexibly adjust the bitrate within a single compact system. The proposed scheme applied to coding depth maps used for view synthesis in a multi-view video coding system. It demonstrates significant bitrate savings on average, maintaining state-of-the-art performance in terms of both PSNR and SSIM indices of rendered views. Withstanding their demonstrated success, the proposed scheme benefits in three major ways: 

\begin{itemize}
\item{Different from systems which generally employ a series of cascaded modules to compress the depth data, our proposed scheme does not introduce cumulative errors because there are few interactions between low-rank modelling and HEVC encoding modules.}
     
\item{The mathematical multi-linear tensor decomposition model for processing depth data is compact and the procedure is not handcrafted, which can efficiently represent the various complex structures in depth maps.}
     
\item{The performance of proposed tensor approximation based scheme is effective for compression with a low bitrate and does not get much
affected with visual artifacts in rendering views from decoded depth maps (e.g., blocky artifacts, blurs, and ringings).
}
\end{itemize}

\section{Related Work} 
Several schemes have been introduced that give natural extension to existing video codec for depth compression \cite{RW1, RW2, RW3, RW4, RW5, RW6}. Belyaev et al. \cite{RW1} proposed method to compress 16 bit depth infrared images via 8 bit depth codecs. They utilize JPEG and H.264/AVC codecs with 8 bits per pixel input format. They compare rate-distortion performance with JPEG 2000, JPEG-XT and H.265/HEVC codecs. Their approach with two 8 bit H.264/AVC codecs achieve similar results as 16 bit HEVC codec. In \cite{RW2,RW3}, Hamout and Elyousfi improved the 3D-HEVC depth map intra prediction model for practical 3D applications. Their approach classified depth video regions as a homogenous or non-homogenous region. Further, they allow conventional intra coding and depth modelling modes step to be skipped. This intuitive approach applies automatic merging possibilistic clustering for region classification based on tensor feature extraction and data analysis. This leads to fast depth map intra coding. Fu et al. \cite{NewRW1} introduced a Depth Intra Skip (DIS) mode that allows early determination for intra mode decision in depth map coding.
Pece et al. \cite{RW4} developed codec for depth streaming by converting single channel depth images of Microsoft Kinect to standard 8 bit three channel images and using existing codecs such as VP8 or H.264. Besides, Pajak et al. \cite{RW5} extends H.264 codec for depth compression by capturing a relation between depth perception and contrast in luminance and improve decompressed visual quality of depth for rendering content. Contrary to above mentioned approaches, Liu et al. \cite{RW6} research suggests that coding scheme based on hybrid lossless-lossy methods provide a better tradeoff between quality and compression ratio for the real-time compression of multiple depth streams. They suggested using x264 for lossy run length encoding to keep the highest bits of 12 bit depth images.

\subsection{Edge-preserving Depth-map Coding} 
Zhang et al.~\cite{Ref1,Ref2} presented a detailed analysis on the computational requirement of depth intra-mode decision of 3D-HEVC. They presented two fast algorithms aiming at speeding up the most time-consuming processes in depth intra-mode. The first technique is based on statistical characteristics of variance distributions in 3D-HEVC depth modelling mode.
The technique introduces an efficient criterion based on the squared Euclidean distance of variances to evaluate rate-distortion costs. Second, a probability-based scheme is proposed to early determine depth intra-mode decision for using segment-wise depth coding based on the low-complexity rate-distortion cost. These proposed approaches provide $33\%–48\%$ time saving with less effect in the coding performance compared with the existing coders.

Kim et al.~\cite{Ref3} represents depth video as a graphical signal. The graph is generated avoiding crossing the depth edges. The scheme employs spectral representation of graph signal and transform kernels to find the eigenvectors of Laplacian matrix of the graph. The scheme requires additional information, \textit{i.e.}, an edge map or an optimal adjacency matrix, into a bitstream for regenerating the exact signal. Their encoder applies context-based adaptive binary arithmetic coding. The scheme is applied to coding depth maps and used for view synthesis in a multi-view video system. Overall, the scheme provides 14\% bit rate savings on average.

Nguyen et al.~\cite{Ref4} proposed techniques to compress a depth video considering coding artifacts, spatial resolution, and dynamic range of the depth data. The coding artifacts around object boundaries are suppressed by weighted mode filtering. The filtering process is also adaptive to reconstruct depth video from the reduced spatial resolution and low dynamic range.
Fabian Jager \cite{Ref5} achieves coding quality by adaptive computation of suitable contour lines, segmenting the depth image. The segments can be approximated with a piecewise-linear function to gain high coding efficiency.
Gao and Smolic \cite{NewRW2} dynamic programming based approach jointly optimize rate-distortion for occlusion-inducing pixels and minimize depth distortion in view synthesis.

Fu et al. \cite{Ref6} explicitly take special characteristics of Kinect-like depth data in coding scheme. Their idea is to first reform depth image by suppressing the depth spatial noises using divisive normalized bilateral filter (DNBL), and then utilize the uniqueness of depth contents for better bit allocation. Their depth measurement error model based on spatial DNBL filtering distinguish the inherent depth edges from the normalized error. This distinction combined with depth padding rebuild the inner depth block continuity and improve efficient block-based coding. The approach saves more than 55\% bitrate with a significant reduction in the coding complexity. Andrew D. Wilson \cite{Ref7} presented a lossless image compression method for 16-bit single channel images typical of Kinect depth cameras. The algorithm is faster than existing lossless techniques. Its performance is demonstrated for 
a network of eight Kinect v2 cameras.

\begin{table*}[t]
\small
\caption{Performance of the HEVC software in coding ``Ballet'' sequence.} 
\centering 

\begin{tabular}{|c|c|c|c|c|c|c|c|}
\hline
     \multicolumn{4}{|c|}{HEVC} & \multicolumn{4}{c|}{HEVC} \\
\hline
\multicolumn{4}{|c|}{Ballet (Camera 3)} & \multicolumn{4}{c|}{Ballet (Camera 5)} \\
\hline
  QP & Bytes & Depth Bitrate (kbps) & Y-PSNR & QP & Bytes & Depth Bitrate (kbps) & Y-PSNR  \\
\hline
 2 & 4200792 &  1680.317 &  62.4805 & 2 & 3978203	& 1591.281 & 62.9252 \\
 \hline
 6 &  2696080 & 1078.432 & 58.4358 & 6 &  2548177 & 1019.271 & 58.8076 \\
\hline
 10 & 1728360 & 691.344 & 55.6816 & 10 & 1618184 & 647.274 & 56.0112 \\
\hline
 14 & 1178418 & 471.367 & 53.5304 & 14  & 1087004 & 434.802 & 53.8004 \\
\hline 
 20 & 703521 & 281.408 & 50.1811 & 20 & 634659 & 253.864 & 50.4475 \\
 \hline 
 26 & 376222 & 150.489 & 45.7444 & 26 & 338075 & 135.23 & 46.0462 \\
\hline 
 38 & 92549 & 37.02 & 36.425 & 38 & 83245 & 33.298 & 36.8471 \\
\hline 
\end{tabular}
\label{TableI}
\end{table*}

\begin{table*}[t]
\small
\caption{Performance of the proposed scheme in coding ``Ballet'' sequence.} 
\centering 

\begin{tabular}{|c|c|c|c|c|c|c|c|c|}
\hline
     \multicolumn{5}{|c|}{Proposed} & \multicolumn{3}{c|}{Proposed} \\
\hline
     \multicolumn{5}{|c|}{Ballet (Camera 3)} & \multicolumn{3}{c|}{Ballet (Camera 5)} \\
\hline
RANK & QP & Bytes & Depth Bitrate (kbps) & Y-PSNR  & Bytes & Depth Bitrate (kbps) & Y-PSNR \\
\hline
RANK 1 & 2 & 2093323	& 837.329 & 62.4673 & 2129651	&  851.86 & 62.3348
\\
\hline
& 6 &  935321 & 374.128 & 59.2998 & 958034 & 383.214 & 59.1788
\\
\hline
& 10 & 475741 & 190.296 & 57.31 & 490424 & 196.17 & 57.2048
\\
\hline
& 14 &  253160 & 101.264 & 55.6207 & 263657 & 105.463 & 55.4672
\\
\hline
& 20 & 107432 & 42.973 & 53.2944 & 112378 & 44.951 & 53.1227
\\
\hline
& 26 & 46086 & 18.434 & 50.4296 & 47853 & 19.141 & 50.2065
\\
\hline
& 38 & 11648 & 4.659 & 43.227 & 12482 & 4.993 & 43.234
\\
\hline
RANK 5 & 2 & 2890680	& 1156.272 & 61.1024 &	2410855 & 964.342 &
62.1882 \\
\hline
& 6 & 1479537  & 591.815	&  58.0589  & 1239664 &	495.866 & 59.0484
\\
\hline
& 10 & 839854 &	 335.942 & 56.0635 & 688043 & 275.217 & 56.9578
\\
\hline
& 14 & 496024 & 198.41 & 54.3667 & 391928 & 156.771 & 55.1635
\\
\hline
& 20 & 235276 &  94.11 &  51.9565 &	176794 & 70.718 & 52.67
\\
\hline
& 26 & 107465 & 42.986  & 48.8446 &	76305 & 30.522 & 49.6478
\\
\hline
& 38 & 26363 & 10.545 & 41.4941 &	17550 & 7.02 & 42.4727
\\
\hline
RANK 10 & 2 & 3494243 & 1397.697 & 60.4022	& 2666111 & 1066.444 & 61.7884
\\
\hline
& 6 & 1915949  & 766.38 & 57.4036	& 1465273 & 586.109 & 58.7244
\\
\hline
& 10 & 1151363	& 460.545 & 55.3769	& 866271 & 346.508 & 56.6165
\\
\hline
& 14 & 721940 &	288.776 & 53.6757  & 526263 & 210.505 & 54.8219
\\
\hline
& 20 & 371092 &	 148.437 &  51.1087 & 259444 &	103.778 & 52.2219
\\
\hline
& 26 &  176086 & 70.434  &  47.626 & 119809 & 47.924 & 48.9273
\\
\hline
& 38 &  44154 &	17.662 & 40.104 & 29028 &	11.611 & 41.4843
\\
\hline
RANK 15 & 2 & 3967986 & 1587.194 & 60.0746	&  3119810 & 1247.924 & 61.2036
\\
\hline
& 6 &  2232791 & 893.116 & 57.0258 & 1755918 & 702.367 & 58.1392
\\
\hline
& 10 & 	1365342 &  546.137 & 54.9812 & 1051988 & 420.795 & 56.0157
\\
\hline
& 14 &  873014 & 349.206 & 	53.2572 & 652055 & 260.822 & 54.223
\\
\hline
& 20 & 454838  & 181.935 & 50.6177	& 325014 & 130.006 & 51.565
\\
\hline
& 26 & 215147 & 86.059 & 47.0147	& 147255 & 58.902 & 48.1186
\\
\hline
& 38 & 50911 &  20.364 & 39.5581	& 33911 & 13.564 & 40.8367
\\
\hline
RANK 20 & 2 & 4503567 & 1801.427 & 59.7568 & 3481704 & 1392.682
& 60.9215
\\
\hline
& 6 & 2593111  & 1037.244	& 56.6565	& 1980722 & 792.289 & 57.8009
\\
\hline
& 10 & 1598926  &	639.57 & 54.5595	& 1193309 & 477.324
& 55.6532
\\
\hline
& 14 &  1026445 & 410.578 &  52.7944 & 745125 & 298.05
& 53.8574
\\
\hline
& 20 & 535315 &  214.126 &  50.0534 & 373015 & 149.206
& 51.1535
\\
\hline
& 26 &  249138 & 99.655 & 46.3509 & 165915 & 66.366
& 47.5866
\\
\hline
& 38 & 55239 &	22.096 &  38.9104 & 36043 & 14.417
& 40.4755
\\
\hline
\end{tabular}
\label{TableII}
\end{table*}

\begin{table*}[t]
\small
\caption{Performance of the HEVC software in coding ``Breakdancing'' sequence.} 
\centering 

\begin{tabular}{|c|c|c|c|c|c|c|c|}
\hline
     \multicolumn{4}{|c|}{HEVC} & \multicolumn{4}{c|}{HEVC} \\
\hline
\multicolumn{4}{|c|}{Breakdancing (Camera 3)} & \multicolumn{4}{c|}{Breakdancing (Camera 5)} \\
\hline
  QP & Bytes & Depth Bitrate (kbps) & Y-PSNR & QP & Bytes & Depth Bitrate (kbps) & Y-PSNR  \\
\hline
 2 &  4926087 & 1970.435 &  62.1439 & 2 & 5071386	& 2028.554 & 62.2019 \\
 \hline
 6 & 3187503 & 1275.001 & 57.7865 & 6 &  3323876 & 1329.55 & 57.7723 \\
\hline
 10 & 2007349 & 802.94 & 54.928 & 10  & 2112154 & 844.862 & 54.8452 \\
\hline
 14 & 1353745  & 541.498 & 52.8257 & 14 & 1427821 & 571.128 & 52.6807 \\
\hline 
 20 & 828067 & 331.227 &  49.6795 & 20 & 870996 & 348.398 & 49.4617 \\
 \hline 
 26 & 428770 & 171.508 &  44.7678 & 26 & 446056 & 178.422 & 44.4987 \\
\hline 
 38 & 81779 & 32.712 & 36.4129 & 38 & 83009 & 33.204 & 36.3369 \\
 
\hline 
\end{tabular}
\label{TableIII}
\end{table*}

\subsection{Segmentation Based Depth Coding} 
Ionut Schiopu and Ioan Tabus \cite{Ref8, Ref9} developed a method for generating sequences of lossy versions of depth image. The sequences are created either by successively merging constant regions of the input depth image, or by iteratively splitting regions from a created lossy depth image employing horizontal or vertical line segments. Their greedy rate-distortion slope optimization algorithms take merge and split decisions greedily, depending on the best slope direction in the rate-distortion curve. They applied suitable entropy coder for compressing these sequences by coding region contours and the optimal depth values of each created lossy image. The obtained results compare favorably over the full range of bitrates with the existing lossy methods. Duch et al. \cite{NewRW3} jointly encode depth and color images in their region-based coding technique. They considered a global 3D scene representation, where segmented color and depth images are organized 
in a coherent hierarchy. This 3D planar decomposition of the scene allows to combine color and depth partitions to obtain the final coding partitions (\textit{i.e.}, segmented regions), without encoding all the depth edges. 
Their rate-distortion methodology demonstrates competitive results with HEVC.
Liu and Kim \cite{NewRW4} followed a quad-tree decomposition approach to partition a depth frame into smooth and edge blocks for encoding.
 
\subsection{Motion Compensation Based Depth Video Coding} 
Wang et al. \cite{Ref10} presented a 3D image warping-based depth video compression algorithm for mobile RGB-D sensors. Their 3D IW-DVC coding framework include motion compensation scheme, designed to exploit the unique characteristics of depth images. The combined egomotion estimation and 3D image warping techniques in their lossless coding scheme adapt depth data with a high dynamic range. 

Zhang et al. \cite{Ref11} depth coding approach analysed high correlation between depth and the corresponding texture video using motion vector and prediction mode. They proposed three efficient, low-complexity approaches based on this correlation for early termination mode decision, adaptive search range motion estimation, and fast disparity estimation. The result outperforms  original 3D-HEVC encoder with reduction about 66\% computational complexity and negligible rate-distortion performance loss.

A number of techniques determine MVs according to the texture video and used this information for depth coding. Shahriyar et al. \cite{IntroRef26} edge-preserving depth map coding scheme used texture MVs to avoid distortion on
the edges. Fan et al. \cite{IntroRef27} proposed a motion estimation method that corrects the depth values in each block using MVs information determined from the color video. Similarly, Lee and Huang \cite{MERef3} extends 2D block matching algorithm with a 3D one, considering horizontal, vertical, and depth dimensions to suppress the coding artifacts. Lei et al. \cite{MERef0} develop a nonsequential coding method for depth maps. Their statistical method suggests that the skip-coding mode and its associated motion vectors in the coded texture can be used for depth coding. It saves bitrate at the cost of a little increase of distortion.

\subsection{Compression Methods Based on Tensor Decomposition}
Most of the compression algorithms based on tensor decomposition rely on
data dependent bases such as factor matrices, instead of choosing the pre-determined ones. Principal component analysis (PCA) as well as Tucker decomposition based approaches adhere to this idea. The Tucker model strives to improve sparsity of transform domain at the cost to reserve the learned bases, which comparatively incline for a three or more dimensions. Some visual data coding algorithms based on Tucker approaches are presented in \cite{TensorNewRef1,TensorNewRef2}. Recently, progressive truncation based approaches for tensor rank reduction demonstrated usefulness in analysing features and structural details in volume data at different scales \cite{TensorNewRef3}. Besides, tensor compression algorithms are explored in the context of 3D displays \cite{TensorNewRef4}, volume rendering and visualization \cite{TensorNewRef5,TensorNewRef6}, and multi-dimensional signal processing applications \cite{TensorNewRef7,TensorNewRef8}. Tensor decompositions and in particular the Tucker model is primarily employed for higher-order compression and dimensionality reduction in the graphics and visualization fields. Ballester-Ripoll and Pajarola \cite{TensorNewRef9} investigated 3D scalar field compression employing Tucker transform coefficients. The study concluded that coefficient thresholding outperforms conventional rank truncation-based methods in terms of quality vs. compression performance. In their algorithm, they combined Tucker core hard thresholding with factor matrix quantization and achieve a better compression rate than slice-wise truncating the core. This approach inspired TAMRESH \cite{TensorNewRef10} and TTHRESH \cite{TensorNewRef9} compression algorithms for multidimensional volume data. TAMRESH large-scale renderer handles input volume by partitioning in small multi-resolution cubic bricks and compressing each brick as a separate higher-order singular value decomposition (HOSVD) core. TTHRESH \cite{TensorNewRef9} performs lossy compression of multidimensional medical data over regular grids. They leverage HOSVD together with the bit-plane, run-length and arithmetic coding in encoding the HOSVD transform coefficients generated in core tensor. The arbitrary target accuracy is supported in their approach via bit-plane coding by greedily compressing bit planes of progressively less importance. The data reduction in their HOSVD-driven approach is achieved by keeping all ranks, however, following attentive lossless compression of all bit planes up to a certain threshold. They also apply this bit-plane based strategy on the factor matrices to encode the data. 

In the last several years, convolutional neural networks (CNNs) based
lossy image compression (LIC) algorithms have been proposed. The end-to-end training in such LIC systems adaptively learns an encoder-decoder pair or adopt specific loss functions to retain image structures and perceptual quality of the decomposed image. Despite such advantages, CNN-based depth video coding algorithms impose challenges. First, CNN-based compressors are only adjusting the bitrate by changing the number of latent feature maps and/or quantized values. The network could be trained specifically for a particular bitrate once at a time. This limits the applicability of CNN-based compressors for practical 3D video rendering systems. Second, updating quantizer in network architecture is hard because of the nondifferentiable property of discrete operation, during the end-to-end training. Cia et al. \cite{TensorNewRef10} CNN-based LIC approach removes these limitations by proposing an effective Tucker Decomposition Network (TDNet), which can adjust multiple bits-per-pixel rates of latent image representation within a single CNN. Their Tucker decomposition layer decomposes a latent image representation into a set of matrices and one small core tensor for lossy coding. However, Cha et al. \cite{TensorNewRef10} TDNet Tucker decomposition layer can not withstand computational complexity of the large amount of 3D video data.

In this paper, we leverage an efficient CANDECOMP/PARAFAC (CP) tensor decomposition algorithm based on alternating least squares (ALS) for depth video coding. Usually, for regular CP-ALS, the cost of decomposition is dominated by the tensor contractions required to solve the quadratic optimization subproblems. Instead, we investigated an efficient adaptation of pairwise perturbation with sparse depth tensors, suggested by Ma and Solomonik \cite{TensorRef1}. The approximation of tensor with factor matrices could be cost effective if perturbative corrections are employed rather than recomputing the tensor contractions to solve the optimization subproblems. In our proposed formulation, we benefit pairwise perturbation to speed up the decomposition procedure as suggested by \cite{TensorRef1}. We observed that approximation to model depth tensor problems with the pairwise perturbation algorithm is accurate as ALS with faster converge to minima with fewer operations.

\section{Proposed Scheme for Depth Video Coding}
An overview of proposed lossy depth video compression scheme is illustrated in Fig.~\ref{Fig1}. The proposed system contains mainly three modules: 1) an encoder module, which consists of two subblocks: first, a tensor decomposition block, which approximates varying low-rank representation of depth video data, and the second HEVC intra encoding subblock, which encodes approximated depth data with varying quantization parameters, 2) a corresponding decoder module, which performs reverse decoding operations, \textit{i.e.}, dequantization and tensor reconstruction, and 3) 3D rendering module, where decoded depth maps are used for novel view synthesis. 

In the first module, the system performs factorization of scene geometry into a set of factor matrices following CP decomposition via alternating least squares algorithm. The representation of the depth data into a high-dimensional sparse tensor domain provides an economic solution to 3D video storage and transmission. Further, performing lossy quantization in the second module with HEVC intra modes on the tensor components of low-rank approximated scene geometry effectively encodes depth data at multiple bitrates. Thus, the scheme overcomes decisive bottlenecks of memory and network bandwidth when handling high-resolution depth frames. The proposed scheme considers a range of rank values for factorizing tensor. Further, encoding with HEVC adopts a variable quantizer to allocate different quantization parameters to the factor matrices. As an indispensable step in multi-view 3D video system, this is a critical step in preserving the major texture and edges in spatially variant depth frames and mitigate rendering artifacts or sampling issues in synthesized views from decoded data. The flexibility proposed scheme offers for changing the rank of factor matrices in tensor decomposition and its quantization levels, facilitates to easily adjust the bitrate and quality levels of  reconstructed depth maps within a compact system. Thus, a single system could enable display adaptation with different 3D devices at reduce storage space, while maintaining backward compatibility and extended functionality such as N-view synthesis without sacrificing much the rendered view quality for baseline adaptation under multiple bitrates.
  
Different components of our proposed depth video compression pipeline are described in the following sections.

\subsection{Tensor Low Rank Approximation}
In module I shown in Fig.~\ref{Fig1}, we performed low-rank approximation of depth video sequences. A CP tensor decomposition model is adopted with an alternating least square procedure \cite{TensorRef1}. The CP tensor decomposition which is a higher-order generalization of the matrix SVD that approximates tensor stack of depth video frames by a sum of $R$ tensor products of vectors. Here, $R$ denotes the decomposition rank. Let us denote tensor created by stacking depth video frames as an order $N$ tensor $\mathfrak{T}_{MES}^{c}$. The CP decomposition is represented by
\begin{equation} 
\mathfrak{T}_{MES}^{c} \approx [[ \textbf{S}^{(1)},  ..., \textbf{S}^{(N)} ]] 
\label{Eq1}
\end{equation} 
where, $\textbf{S}^{(i)} = [ s^{(i)}_{1}, ..., s^{(i)}_{r}]$. Writing (\ref{Eq1}) by a sum of $R$ tensor products of vectors serves to approximate $\mathfrak{T}_{MES}^{c}$ as
\begin{equation}
\mathfrak{T}_{MES}^{c} \approx \sum_{r=1}^{R} s^{(1)}_{r} \circ \cdot \cdot \cdot \circ s^{(N)}_{r}
\end{equation} 
The conventional CP-ALS alternates among quadratic optimization problems for each of the factor matrices $\textbf{S}^{(n)}$. This requires to solve linear least squares problems for each row,
\begin{equation}
\textbf{S}_{new}^{(n)} \textbf{K}^{(n)T} \cong
\mathfrak{\textbf{X}}_{(n)}
\label{Eq3} 
\end{equation} 
The matrix $\textbf{K}^{(n)} \in \mathfrak{R}^{I_n \times R}$ is formed by Khatri-Rao products of the other factor matrices
\begin{equation}
\textbf{K}^{(n)} = \textbf{S}^{(1)} \odot \cdot \cdot \cdot \odot \textbf{S}^{(n-1)} \odot \textbf{S}^{(n+1)} \odot \cdot \cdot \cdot 
\textbf{S}^{(N)}
\label{Eq4}
\end{equation} 
where, $I_n = a_1 \times \cdot \cdot \cdot \times a_{n-1} \times a_{n+1} \times \cdot \cdot \cdot a_{N}$. The linear least squares problems (\ref{Eq3}) are usually solved following the pairwise perturbation method with normal equations 
\begin{equation}
\textbf{S}_{new}^{(n)} {\Gamma}^{(n)} = 
\mathfrak{\textbf{X}}_{(n)} \textbf{K}^{(n)}
\label{Eq5}
\end{equation} 
and computing $\Gamma \in \mathfrak{R}^{R \times R}$ as 
\begin{equation}
\Gamma^{(n)} = 
\textbf{A}^{(1)} \ast \cdot \cdot \cdot \ast  \textbf{A}^{(n-1)} \ast  \textbf{A}^{(n+1)} \ast \cdot \cdot \cdot 
\textbf{A}^{(N)}
\end{equation} 
with each $\textbf{A}^{(i)} = \textbf{S}^{(i)T} \textbf{S}^{(i)}$. 
The equations (\ref{Eq5}) could be solved by formulating the problem as an unconstrained minimization of the nonlinear objective function,
\begin{equation}
f(\textbf{S}^{(1)},  ..., \textbf{S}^{(N)}) = \frac{1}{2} || \mathfrak{T}_{MES}^{c} - [[ \textbf{S}^{(1)},  ..., \textbf{S}^{(N)} ]] ||^{F}_{2}
\end{equation} 
with $n^{th}$ gradient component computed as
\begin{equation}
\frac{\partial f}{\partial \textbf{S}^{(n)}} = \textbf{S}^{(n)} {\Gamma}^{(n)} - \mathfrak{\textbf{X}}_{(n)}\textbf{K}^{(n)} = (\textbf{S}^{(n)} - \textbf{S}_{new}^{(n)}){\Gamma}^{(n)} 
\end{equation} 
This CP decomposition works if ascertain convergence of Frobenius norm of the $N$ components of the overall gradient. To fasten the computation, we adopted Ma et al. \cite{TensorRef2} pairwise perturbation approach which overcomes the  main computational bottleneck of CP-ALS caused by Matricized Tensor Times Khatri-Rao Product $  \textbf{M}^{n}=\mathfrak{\textbf{X}}_{(n)} \textbf{K}^{(n)}$ estimation. The ALS procedure for CP decomposition is accelerated to approximate $\textbf{M}^{n} \approx \tilde{\textbf{M}}^{n}$. Let $\textbf{M}^{n}$ can be expressed as
\begin{equation}
\textbf{M}^{n} = \mathfrak{\textbf{X}}_{(n)} \bigodot_{i=1,i \neq n}^{N} (\textbf{S}_{p}^{(i)} + d\textbf{S}^{(i)})
\end{equation} 
where, $\textbf{S}_{p}^{(n)}$ denotes the $\textbf{S}^{(n)}$ computed with a standard ALS step at some (\textit{i.e.}, $p$) number of steps preceding to the present one.
Thus, $\textbf{S}^{(n)}$ at the current step can be expressed as
\begin{equation} 
\textbf{S}^{(n)} = \textbf{S}_{p}^{(n)} + d\textbf{S}^{(n)}
\end{equation} 
The pairwise perturbation algorithm of Ma et al. \cite{TensorRef1} effectively approximates
$\textbf{M}^{n}$ as 
\begin{equation} 
\tilde{\textbf{M}}^{n}(y,k) = \textbf{M}_{p}^{(n)}(y,k) + \sum_{i=1,i \neq n}^{N} \sum_{x=1}^{a_i} \mathfrak{M}_{p}^{(i,n)}(x,y,k) d\textbf{S}^{i}(x,k)
\end{equation} 
where,
\begin{equation}
\textbf{M}_{p}^{n} = \mathfrak{\textbf{X}}_{(n)} \bigodot_{i=1,i \neq n}^{N} \textbf{S}_{p}^{(i)}
\end{equation} 
and 
\begin{equation}
\mathfrak{M}_{p}^{(i,n)} = \mathfrak{T}_{MES_{(i,n)}}^{c}  
\bigodot_{j=\{1,...,N\} \setminus \{i,n\}}^{N} \textbf{S}_{p}^{(j)}
\end{equation} 
The $\mathfrak{M}_{p}^{(i,n)}$ is defined as $\mathfrak{T}_{MES}^{c}$ contracted with
$\textbf{S}_{p}^{(j)}$ for $i \in \{1,...,N\} \setminus \{i,n\}$. Given $\textbf{M}_{p}^{(n)}$ and $\mathfrak{M}_{p}^{(i,n)}$, the $\tilde{\textbf{M}}^{(n)}$ is computed for all $n \in \{1,...,N\}$ efficiently in small number of operations. We adopted dimension trees based approach \cite{TensorRef2,TensorRef3} for the computation of pairwise perturbation operators $\textbf{M}_{p}^{(n)}$ and $\mathfrak{M}_{p}^{(i,n)}$.

\subsection{HEVC Coding for Low Rank Depth Video}
We encode the low-rank approximated tensor components of depth video (\textit{i.e.} factor matrices) via HEVC intra coding tools in module II, as shown in Fig.~\ref{Fig2}. High Efficiency Video Coding (HEVC) compression standard is designed under the collaborative standardization project of ITU-T VCEG and ISO/IEC MPEG \cite{HEVCRef1}. It is developed as part of the MPEG-H project, which is a successor of widely used Advanced Video Coding (H.264, or MPEG-4 Part 10) standard. The detail description of the changes relative to H.264/MPEG-4 AVC is given in \cite{HEVCRef2}. 

In HEVC, coding tree blocks (CTBs) based partition approach is adopted in encoding a view. The encoder selects the size of CTBs according to architectural characteristics and application requirements such as delay and memory constraints. A coding tree unit (CTU) processes the luma CTB and the two chroma CTBs. Typically, $N \times N$ samples of the luma component and 
the corresponding $(N/2) \times (N/2)$ samples of two chroma components are signaled inside the bitstream. The CTU is defined as the basic processing unit, which identify the decoding process in the standard. The luma and chroma blocks in CTBs can be partitioned further into multiple coding blocks (CBs). The quadtree syntax of CTU allows splitting into variable size blocks taken into account the attributes of the region covered by the CTB. The CB size is specified by the decoder syntax. It could be of minimum size $8 \times 8$
for luma samples or larger. For a coding unit (CU), where luma and the chroma CBs are processed, a prediction mode is signaled. The prediction mode can
be either an intra or inter mode. In intra prediction, thirty five prediction modes are specified for the luma CB. A single intra prediction mode is signaled for both chroma CBs. The same prediction mode is specified for luma or a horizontal, vertical, planar, left-downward diagonal or DCT (discrete cosine transform) prediction mode \cite{HEVCRef3}. The intra mode is applied independently for each transform block. In the inter-mode coding units, the luma and chroma coding blocks are associated with one, two, or four luma and chroma prediction blocks. The inter-mode determine one or two motion vectors for each prediction unit following unipredictive or bipredictive coding. Asymmetric motion partitioning and splitting is applied appropriately to chroma coding blocks.

HEVC supports several common features related to H.264/MPEG-4 AVC. It facilities quarter sample precision motion vectors, weighted prediction and multiple reference pictures. The concepts of I, P, and B slices are similar to H.264/MPEG-4 AVC. We used latest HEVC reference software HM-16.6 in the proposed formulation. The advanced HEVC version supports so called merge mode, which substantially improved the coding of motion parameters compared to earlier standards. In merge mode, no motion parameters are coded. Alternatively, a candidate list of motion parameters are derived which includes motion parameters of spatially neighboring blocks and temporally predicted motion parameters for the corresponding prediction unit. This information is derived considering motion data of a co-located block in a reference image. 
This advanced mode for motion compensation addresses the challenges of large block sizes and consistently displaced image regions, \textit{e.g.}, due to the object's motion. 

HEVC supports residual quadtree (RQT) split of luma and chroma CBs for coding the inter or intra prediction residual signal. In RQT, either CB is represented as sole luma transform block or four equal-sized luma transform blocks. In split mode of RQT, each resulting luma transform block can be further broken into four smaller luma transform blocks. Similarly, split applies to the chroma CB. The associated syntax of RQT with the luma and chroma transform blocks form a transform unit (TU) where a 2D transform (\textit{i.e.} DCT approximation) is applied to luma and chroma samples.

Besides, HEVC like H.264/MPEG-4 AVC supports entropy coded using CABAC, 
improved sophisticated context selection scheme for transform coefficient
coding, and sample-adaptive coding for efficient motion compensation that
reduces the distortion in encoding and decoding samples. The HEVC Main profile (MP) supports coding of 8-bit-per-sample video. Two additional profiles are specified in the later HEVC Drafts 9-10 specification: the one Main 10 profile (for 10-bit-per-sample video) and the second Main Still Picture profile for coding still images employing only intra-coding tools.

\subsection{View Synthesis}
In module III shown in Fig.~\ref{Fig1}, we performed depth-image-based free-viewpoint rendering using the decoded depth map of the left and right
cameras. DIBR method proposed by Sharma et al.~\cite{IntroRef9} is used for virtual view synthesis. The imperfections of depth images cause unrealistic artifacts and affect rendered view quality in DIBR systems. Thus, it is more appropriate to consider the effect of decoded depth maps on synthesized view quality, instead of observing the distortion in compressed depth data. The PSNR and SSIM metrics are used to estimate the rendered view distortion caused by the errors in decoded depth images compressed using the proposed coding scheme and HEVC codec. For a fair evaluation of depth coding, we have not used the captured original reference video at the virtual camera position to measure the quality. This is to prevent distortion incorporated in quality measurement caused by camera parameters and lighting variation among multi-view video data. PSNR and SSIM scores are computed by comparing the synthesized video without compression of depth videos as the reference and the synthesized video using the decoded depth videos \cite{EvalRender3d}. 

\section{Experimental Results}
A series of experiments are conducted to assess the performance of the proposed depth compression scheme. We tested our model on ``Ballet'' and ``Breakdancing'' sequences provided by Microsoft Research. The multi-camera data is acquired by eight cameras located along a 1D arc covering about $30^{\circ}$ from one end to the other. The color and depth videos are provided with resolutions of $1024 \times 768$ \cite{MSRdata}.

We encode depth video of each camera sequence using the proposed coding model and High Efficiency Video Coding (HEVC) reference software 
HM-16.0 \cite{HEVCsoftware}. The color and depth video of left and right cameras are encoded considering various quantization parameters (QPs). The QP values of $2$, $6$, $10$, $14$, $20$, $26$, $38$ are used to encode the depth videos.

The experimental model to test the performance of proposed 
scheme analysed the quality of synthesized intermediate view. 
We selected reference cameras $3^{rd}$ and $5^{th}$  among eight cameras of ``Ballet'' and ``Breakdancing'' multi-camera data. The intermediate views are synthesized at virtual camera $4^{th}$ using free-viewpoint rendering algorithm developed by Sharma et al.~\cite{IntroRef9}. The objective quality assessment is performed by measuring Peak Signal-to-Noise ratio (PSNR) and structural similarity index measure (SSIM). The PSNR is measured by comparing the synthesized video rendered using original depth maps as the reference and the synthesized video using the decoded depth maps. It is critical to note that decoded video is used for virtual view synthesis to measure the distortion caused by depth map compression. The rate distortion (RD) curves are plotted considering the total bitrate required to encode depth video of both reference cameras and PSNR/SSIM scores of the synthesized virtual video. 

Fig.~\ref{Fig2} shows the ``Bitrate vs PSNR'' and ``Bitrate vs SSIM'' comparison graphs between the proposed method and HEVC codec. The x-axis plots total bitrate required to encode left (camera 3) and right camera (camera 5) depth sequence. The y-axis plots PSNR/SSIM measures of the synthesized intermediate camera view (camera 4). We achieve significant bitrate reduction using proposed coding scheme compared to directly encode the depth video of left and right camera using HEVC codec, while maintaining appropriate quality of synthesized virtual camera views in terms of PSNR and SSIM measures. The encoder results obtained by applying proposed scheme and HEVC in compressing ``Ballet'' and ``Breakdancing'' depth sequences are tabulated in Table~\ref{TableI}-\ref{TableIV}. Total number of bytes required to encode left (camera 3) and right (camera 5) depth sequences are summarized in Table~\ref{TableVI}-\ref{TableVIII}. The Bjontegaard metric calculation results demonstrate the coding efficiency of proposed scheme compared to HEVC codec \cite{BDBRRef}. The Bjontegaard delta bitrate (BDBR) reduction results considering different ranks and quantization values are given in Table~\ref{TableV}. The BDBR measures are computed by considering the total bitrate needed to encode left and right camera depth video and PSNR scores of the synthesized intermediate camera video. Note that minus sign indicates bitrate reduction. Compared to HEVC, our scheme achieves 62.8143\%, 35.7449\%, 43.1331\%, 21.5080\%, 30.7221\% rate gains on average with ``Ballet'' data, considering tensor ranks $1$, $5$, $10$, $15$, $20$ respectively. On ``Breakdancing'' data, we obtain 70.7924\%, 73.3179\%, 49.1670\%, 26.6356\%, 16.2384\% rate gains, considering tensor ranks $1$, $5$, $10$, $15$, $20$ respectively. This is a substantial improvement over HEVC for directly encoding depth videos with the same quantization parameters and configuration. It can also be observed in Fig~\ref{Fig2} that for higher ranks, particularly large PSNR gains are obtained using the proposed model even for low bitrates, comparable to HEVC. The SSIM scores demonstrate that the perceived quality of synthesized views is also appropriate for different bitrates, obtained considering varying tensor decomposition ranks and quantization parameters in our scheme. This explains significant redundancies can be removed, which lowers the overhead in transmission for bitrate coding without affecting much the perceived change in structural information of rendered views quality. Another critical advantage of our scheme is to allow scalable depth video coding. Since the proposed scheme can render virtual views of varying quality at the decoder considering different ranks and QP values. This flexibility could be preferable in 3D video transmission and broadcasting scenarios as it provides additional levels of scalability.

\begin{table*}[t]
\small
\caption{Performance of the proposed scheme in coding ``Breakdancing'' sequence.} 
\centering 

\begin{tabular}{|c|c|c|c|c|c|c|c|c|}
\hline
     \multicolumn{5}{|c|}{Proposed} & \multicolumn{3}{c|}{Proposed} \\
\hline
     \multicolumn{5}{|c|}{Breakdancing (Camera 3)} & \multicolumn{3}{c|}{Breakdancing (Camera 5)} \\
\hline
RANK & QP & Bytes & Depth Bitrate (kbps) & Y-PSNR  & Bytes & Depth Bitrate (kbps) & Y-PSNR \\
\hline
RANK 1 & 2 & 2367817	& 947.127 &  61.8849 & 	2351345
& 940.538 & 61.7692
\\
\hline
& 6 & 1130354 & 452.142 &  58.8111 & 1133670
& 453.468 & 58.7342
\\
\hline
& 10 & 607790 & 243.116 & 56.8681 & 613464
& 245.386 & 56.8112
\\
\hline
& 14 &  341031 & 136.412 & 55.2676 & 347297
& 138.919 & 55.2019
\\
\hline
& 20 & 156412 &  62.565 & 53.0796 & 160576
& 64.23 & 53.0383
\\
\hline
& 26 & 70544 & 28.218 &  50.1954 & 73637
& 29.455 & 50.244
\\
\hline
& 38 & 19246 & 7.698 & 42.2817 & 20801
& 8.32 & 42.5151
\\
\hline
RANK 5 & 2 & 2868655 & 1147.462 & 60.6944 &	2921655 & 1168.662 & 60.6136
\\
\hline
& 6 & 1489113  & 595.645 & 57.8249 & 1512204 &	604.882 & 57.7453
\\
\hline
& 10 & 874402 &	349.761 &  55.8803 & 890661 & 356.264 & 55.8043
\\
\hline
& 14 & 534396 & 213.758 & 54.2632	& 542109 & 216.844 & 54.1774
\\
\hline
& 20 & 266753 & 106.701 & 51.9505 &	268900 & 107.56 & 51.8714
\\
\hline
& 26 & 129442 &  51.777 & 48.7986 &	128559 & 51.424 & 48.7618
\\
\hline 
& 38 & 35465 &  14.186 & 41.1126 &	34696 &  13.878 & 41.2135
\\
\hline
RANK 10 & 2 & 3424073	&  1369.629 & 59.9562	& 3499848
& 1399.939 & 59.8421
\\
\hline
& 6 & 1859422  & 743.769 & 	57.1562 & 1912219 & 764.888 & 57.0479
\\
\hline
& 10 & 	1150623 & 460.249  & 55.2579	& 1194034 & 477.614 & 55.1332
\\
\hline
& 14 & 731822  & 292.729	&  53.6425 & 760978 & 304.391 & 53.4965
\\
\hline
& 20 & 379396 &	 151.758 & 51.1748 & 393172 &	157.269  & 51.0177
\\
\hline
& 26 &  186746 & 74.698	 &  47.7901 & 190911 & 76.364 & 47.6202
\\
\hline
& 38 &  50172 &	20.069 & 40.1731  & 50211 &	20.084 & 40.1083
\\
\hline
RANK 15 & 2 & 3790115	& 1516.046 & 	59.6792 & 3939703
& 1575.881 & 59.532
\\
\hline
& 6 & 2104438  &	841.775  & 56.8484  & 2197956
& 879.182
& 56.6962
\\
\hline
& 10 & 	1309412 &  523.765 & 54.9319 & 1382325
& 552.93
& 54.7757
\\
\hline
& 14 & 841605 & 336.642 & 53.3062	& 891563
& 356.625
& 53.137
\\
\hline
& 20 & 438737  & 175.495	& 	50.7645 & 463942
& 185.577
& 50.5765
\\
\hline
& 26 & 212934 &  85.174 & 47.2708	& 223156
& 89.262
& 47.0653
\\
\hline
& 38 &  54525 &  21.81 & 39.648	& 56255
& 22.502
& 39.5446
\\
\hline
RANK 20 & 2 & 4143552 & 1657.421 & 59.4377 & 4306444
& 1722.578
& 59.2869
\\
\hline
& 6 & 2331086 & 932.434	& 56.5539	& 2445676
& 978.27
& 56.4193
\\
\hline
& 10 &  1462222 &	584.889 & 54.6463	& 1544937
& 617.975
 & 54.4916
\\
\hline
& 14 & 943868 &  377.547 & 53.0033 & 1001271
& 400.508
& 52.8347
\\
\hline
& 20 & 493315 &  197.326 &  50.4118 & 522700
& 209.08
& 50.2065
\\
\hline
& 26 &  236580 &  94.632 & 46.8621 & 248178
& 99.271
& 46.6374
\\
\hline
& 38 & 58277 & 23.311 &  39.3249 & 60115
& 24.046
& 39.1823
\\
\hline
\end{tabular}
\label{TableIV}
\end{table*}

\begin{table}[t]
\small
\caption{Bjontegaard metric calculation results compared to HEVC codec.} 
\centering 

\begin{tabular}{|c|c|c|c|c|c|c|c|}
\hline
 &  &  Ballet & Breakdancing \\
\hline
RANK & QP  & BDBR (\%) & BDBR (\%) \\
\hline
RANK 1 & 2 &  -91.7644 & -91.4810
\\
\hline
& 6 & -87.563 & -88.1485
\\
\hline
& 10 	& -82.5545 & -84.3392
\\
\hline
& 14   &  -74.6781 & -78.3987
\\
\hline
& 20  & -56.571 & -64.9441
\\
\hline
& 26     &  -28.9834 &  -49.3263
\\
\hline
& 38     & -17.586 &  -38.9087
\\
\hline
Average &     & \textbf{-62.8143}    &  \textbf{ -70.7924}
\\
\hline
 &      &  & 
\\
\hline
RANK 5 & 2  	&  -61.2031 & -87.2847
\\
\hline
& 6    & 	-50.9729 & -83.9632
\\
\hline
& 10    &  -42.1304 &  -81.3858
\\
\hline
& 14  &	-29.3045 &  -77.3789
\\
\hline
& 20    & -2.9364 &  -68.0988
\\
\hline
& 26     & -22.4574 &  -59.3702
\\
\hline
& 38     & -41.2093 &  -55.7434
\\
\hline
Average &      & \textbf{-35.7449} &  \textbf{ -73.3179}
\\
\hline
 &      &  & 
\\
\hline
RANK 10 & 2 	&  -66.6627 & -66.2222
\\
\hline
& 6   & -60.9403 & -59.6894
\\
\hline
& 10	 & -56.9193 & -57.4577
\\
\hline
& 14   &  -50.9719 &  -52.7528
\\
\hline
& 20   &  -38.9249 &   -40.1808
\\
\hline
& 26   & -23.6252 &  -30.8459
\\
\hline
& 38   &  -3.8872 & -37.0202
\\
\hline
Average &      &  \textbf{-43.1331} & \textbf{-49.1670}
\\
\hline
 &      &  & 
\\
\hline

RANK 15 & 2 & 	-37.6335 &  -46.5526
\\
\hline
& 6 &  -28.1844 &  -37.5994
\\
\hline
& 10	& -22.922 &   -35.5160
\\
\hline
& 14  &  -15.2211 &   -30.5218
\\
\hline
& 20  	& 	-1.1148 &   -15.0541
\\
\hline
& 26  & -17.0494 &  -4.1785
\\
\hline
& 38    & 	-28.4309 &   -17.0270
\\
\hline
Average &      &   \textbf{-21.5080} & \textbf{ -26.6356}
\\
\hline
 &      &  & 
\\
\hline

RANK 20 & 2 & -49.8622 & -33.7813
\\
\hline
& 6 &  -43.92 & -23.8671
\\
\hline
& 10   & 	-40.5368 &  -22.4193
\\
\hline
& 14   &  -35.388 &  -17.5907
\\
\hline
& 20  &  -23.2538 & -1.0055
\\
\hline
& 26    & -5.8719 &  -9.8350
\\
\hline
& 38 	& -16.2222 &  -5.1702
\\
\hline
Average &      & \textbf{-30.7221} &  \textbf{ -16.2384}
\\
\hline
\end{tabular}
\label{TableV}
\end{table}

\begin{table}[t]
\small
\caption{Performance of the HEVC codec on MSR 3D video sequences. Total number of bytes required to encode left (camera 3) and right (camera 5) depth sequences.} 
\centering

\begin{tabular}{|c|c|c|p{1.5cm}|p{1.5cm}|c|c|c|c|c|}
\hline
    SCENE & \textbf{QP}  & \multicolumn{3}{|c|}{HEVC Codec}  \\
\hline
& & Camera 3 & Camera 5 & Total Bytes  \\
\hline
Ballet & 2 &	4200792	& 3978203	& 8178995
\\
\hline
& 6	&	2696080	& 2548177	&  5244257
\\
\hline
& 10 	&	1728360	& 1618184 & 3346544
\\
\hline
& 14 	&	1178418	& 1087004  & 2265422
\\
\hline
& 20 	&	703521	& 634659	&	1338180
\\
\hline
& 26 	&	376222	& 338075	&	 714297
\\
\hline
& 38 	& 92549 &	83245 &	 175794 \\
\hline
Breakdancing & 2 & 4926087 &	 5071386 &	9997473
\\
\hline
& 6	&	3187503	& 3323876 	&  6511379
\\
\hline
& 10 	&	2007349	& 2112154	&	4119503
\\
\hline
& 14 	&	1353745	& 1427821	&	  2781566
\\
\hline
& 20 	&	828067	& 870996	&	 1699063
\\
\hline
& 26 	&	428770	& 446056	 & 874826
\\
\hline
& 38 	& 81779 &	83009 &	164788 \\
\hline
\end{tabular}    
    
\label{TableVI}
\end{table}

\begin{table}[t]
\small
\caption{Performance of the proposed scheme on ``Ballet'' sequence.  Total number of bytes required to encode left (camera 3) and right (camera 5) depth sequences.} 
\centering

\begin{tabular}{|c|c|c|c|c|c|c|c|}
\hline
     \multicolumn{5}{|c|}{} \\
\hline
RANK & QP & Camera 3 & Camera 5 & Total Bytes \\
\hline
RANK 1 & 2 & 2093323	& 2129651	& 4222974
\\
\hline
& 6 & 	935321	& 958034 &	1893355	\\
\hline
& 10 & 	475741 & 490424 &	966165
\\
\hline
& 14 & 	253160 &	263657 & 516817
\\
\hline
& 20 & 	107432 &	112378 &	219810
\\
\hline
& 26 & 	 46086 &	47853 & 93939
\\
\hline
& 38 & 	11648 &	12482 & 24130 \\
\hline
RANK 5 & 2 & 2890680 & 2410855 &	5301535
\\
\hline
& 6 & 	1479537	& 1239664 &	2719201	\\
\hline
& 10 & 	839854 & 688043 & 1527897
\\
\hline
& 14 & 	496024 &	391928 & 887952
\\
\hline
& 20 & 	235276 &	176794 & 412070
\\
\hline
& 26 & 	 107465 &	76305 & 183770
\\
\hline
& 38 & 	26363 &	17550 & 43913 \\	
\hline
\hline
RANK 10 & 2 & 3494243	& 2666111 &	6160354
\\
\hline
& 6 & 	1915949 & 1465273 &	3381222
	\\
	\hline
& 10 & 1151363	& 866271 &	2017634
\\
\hline
& 14 & 	721940 &	526263 & 1248203
\\
\hline
& 20 & 	 371092 &	259444 &	630536
\\
\hline
& 26 & 	176086 &	119809 & 295895
\\
\hline
\hline
& 38 & 	44154 &	29028 & 73182\\
\hline
RANK 15 & 2 & 3967986 &	3119810 & 7087796
\\
\hline
& 6 & 	2232791	& 1755918 &	3988709 \\
\hline
& 10 & 1365342	& 1051988 &	2417330
\\
\hline
& 14 & 	873014 &	652055 & 1525069
\\
\hline
& 20 & 	454838 &	325014 &	779852
\\
\hline
& 26 & 	215147 &	147255 & 362402
\\
\hline
& 38 & 	50911 &	33911 & 84822 \\
\hline
RANK 20 & 2 &	4503567 & 3481704 &	7985271
\\
\hline
& 6 & 	2593111	& 1980722 &	4573833
	\\
\hline
& 10 & 	1598926 & 1193309 &	2792235
\\
\hline
& 14 & 	1026445 &	745125 & 1771570
\\
\hline
& 20 & 	535315 &	373015 &	908330
\\
\hline
& 26 & 	249138 &	165915 & 415053
\\
\hline
& 38 & 	 55239 &	36043 & 91282 \\
\hline
\end{tabular}    
\label{TableVII}
\end{table}

\begin{table}[t]
\small
\caption{Performance of the proposed scheme on ``Breakdancing'' sequence. Total number of bytes required to encode left (camera 3) and right (camera 5) depth sequences.} 
\centering

\begin{tabular}{|c|c|c|c|c|c|c|c|}
\hline
     \multicolumn{5}{|c|}{} \\
\hline
RANK & QP & Camera 3 & Camera 5 & Total Bytes \\
\hline
RANK 1 & 2 & 2367817	& 2351345 & 4719162
\\
\hline
& 6 & 1130354		& 1133670 &	2264024 \\
\hline
& 10 & 	607790 & 613464 &	1221254
\\
\hline
& 14 & 341031	 &	347297 & 688328
\\
\hline
& 20 & 	156412 &	160576 &	316988
\\
\hline
& 26 & 	70544 &	73637 & 144181
\\
\hline
& 38 & 19246	 &	20801 & 40047 \\
\hline
RANK 5 & 2 & 2868655 & 2921655 &	5790310
\\
\hline
& 6 & 	1489113	& 1512204 &	3001317 
\\
\hline
& 10 & 	874402 & 890661 &	1765063
\\
\hline
& 14 & 	534396 &	542109 & 1076505
\\
\hline
& 20 & 	266753 &	268900 &	535653
\\
\hline
& 26 & 	129442 &	128559 & 258001
\\
\hline
& 38 & 	35465 &	34696 & 70161 \\
\hline
RANK 10 & 2 & 3424073	& 3499848 &	6923921
\\
\hline
& 6 & 1859422		& 1912219 &	3771641 \\
\hline
& 10 & 1150623	& 1194034 &	2344657
\\
\hline
& 14 & 	731822 &	760978 & 1492800
\\
\hline
& 20 & 	379396 &	393172 &	772568
\\
\hline
& 26 & 	186746 &	190911 & 377657
\\
\hline
& 38 & 	50172 &	50211 & 100383 \\
\hline
RANK 15 & 2 & 3790115 &	3939703 & 7729818
\\
\hline
& 6 & 	2104438	& 2197956 &	4302394 \\
\hline
& 10 & 1309412 & 1382325 & 2691737
\\
\hline
& 14 & 	841605 &	891563 & 1733168
\\
\hline
& 20 & 	438737 &	463942 &	902679
\\
\hline
& 26 & 	212934 &	223156 & 436090
\\
\hline
& 38 & 	54525 &	56255 & 110780
\\
\hline
RANK 20 & 2 & 4143552	& 4306444 &	8449996 
\\
\hline
& 6 &  2331086	& 2445676 &	4776762	\\
\hline
& 10 & 1462222 &	1544937 & 3007159
\\
\hline
& 14 & 	943868 & 1001271 & 1945139 \\
\hline
& 20 & 	493315 &	522700 &	1016015
\\
\hline
& 26 & 	236580 &	248178 & 484758
\\
\hline
& 38 & 	58277 &	60115 & 118392 
\\
\hline
\end{tabular}    
\label{TableVIII}
\end{table}

\section{Conclusion}
In this paper, we presented a simple yet effective depth video codec as an extension of the HEVC standard for stereoscopic and autostereoscopic multi-view displays. The proposed mathematical scheme is developed on tensor modelling of higher-order information in visual depth data. Our system decomposes a tensor representation of scene geometry into a set of factor matrices following CP decomposition via alternating least squares for lossy depth compression. By varying the rank of factor matrices and its quantization levels, we could smoothly adapt the bitrate in coding the depth content. Thus, appropriately achieved the goal of using a single system to cover a range of bit rates and quality levels. The scheme offers scalable and flexible encoder for coding of depth content, its delivery and display on a variety of 3D device in terms of supported spatial and temporal resolution and decoder complexity. 

Our depth coding scheme is applicable to be used together with other hybrid  coding architectures such as, \textit{e.g.}, MPEG-4 Visual, H.264/MPEG-4 AVC, MV-HEVC, 3D-HEVC, H.263, and H.262/MPEG-2, etc. Low-rank decomposition by pairwise perturbation retain the multi-dimensional dependencies in activation tensors of depth data and results in significant bitrate savings. This encoder control of the depth data and view synthesis quality, guarantees display adaptation via intermediate view generation at the decoder. Using a subset of the coded depth components, 2D video, stereo video or full MVD can be reconstructed from the 3D decoded depth bitstreams and reference texture data. The objective results shown in Table~\ref{TableV} show significant bitrate reduction (BDBR) in comparison to HEVC, while retaining the almost similar synthesis quality. The PSNR and SSIM measurements demonstrate satisfactory performance and not much obtrusion due to geometrical distortions in the synthesis coming from compression along depth discontinuities.
In the future, we analyse view quality considering different 
factors such as camera baseline variation, scene characteristics, artifacts
caused by inherent visibility, disocclusion, and resampling problems in DIBR \cite{IntroRef9,IntroMy1,IntroMy2,IntroMy3}. This enables a more precise estimate of synthesize view quality which helps to gain an optimal RD performance, choosing different parameters and mode selection in low-rank tensor based coding algorithms for 3D display applications.
  
Other avenues for future research include exploration of different tensor models in revealing latent correlations residing in high dimensional spaces of 3D video data and remove redundancies using optimized linear/multilinear algebra. A generic higher-order tensor representation in the context of an efficient HEVC extension for MVD coding is critical. While matrices based coding methods are limited to a single mode of variation, however, to naturally accommodate different modes of variation, tensors could be used in MVD video compression. We would develop an adaptive control mechanism that would allow for efficient representation of the MVD content description along each mode by a factor matrix and analyze different grounds of the decomposition with modes interaction by a core coefficient tensor for model reduction applications. We believe preliminary coding results with CP decomposition reported in this paper could motivate further investigation of tensor-based methods as a crucial mathematical object for representation and standardization of improved coding tools for candidate extensions of HEVC and its variant for 3D display applications. This could particularly benefit for further space savings for low bit rates, higher-resolution 3D content and low-delay application encodings. We would also explore tensor regression approaches for achieving higher coding gains.

\bibliographystyle{ieeetr}
\footnotesize
\bibliography{root.bib}
\end{document}